\documentstyle[12pt]{article}
\newcommand{\ssigma}{\hbox{$\kern2.5pt\vrule height4pt\kern-2.5pt\sigma$}}
\newcommand{\oone}{\hbox{$1\kern-3pt\hbox{\rm l}$}}

\newcommand{\GeV}{{\sl\,GeV}}
\newcommand{\real}{{\sl Re\,}}

\newcommand{\GG}{{\cal G}}
\newcommand{\GL}{{\cal G}_\ell}
\newcommand{\Li}{{\rm Li}_2}

\begin{document}
\thispagestyle{empty}
\begin{flushright}
MZ-TH/96-34\\
hep-ph/9704416\\
July 1997\\
%(Version of~\today)\\
\end{flushright}
\vspace{0.5cm}
\begin{center}
{\Large\bf Gluon Polarization in \boldmath{$e^+e^-\rightarrow t\bar tG$}}\\
\vspace{1.3cm}
{\large S.~Groote, J.G.~K\"orner and J.A.~Leyva\footnote{On leave of absence 
from CIF, Bogot\'a, Colombia}}\\[1cm]
Institut f\"ur Physik, Johannes-Gutenberg-Universit\"at,\\[.2cm]
Staudinger Weg 7, D-55099 Mainz, Germany\\
\end{center}
\vspace{1cm}

\begin{abstract}\noindent
We calculate the linear polarization of gluons radiated off top quarks 
produced in $e^+e^-$ annihilations. For typical top pair production energies 
at the Next-Linear-Collider (NLC) the degree of linear polarization remains 
close to its soft gluon value of 100\% over almost the whole energy spectrum 
of the gluon. The massive quark results are compared with the corresponding 
results for the massless quark case.
\end{abstract}

\newpage

\section{Introduction}
The polarization of gluons in $e^+e^-$ annihilation~\cite{gluon1,gluon2}, 
in deep inelastic scattering~\cite{gluon3} and in quarkonium 
decays~\cite{gluon2,gluon4} has been studied in a series of papers dating 
back to the early 80's. Several proposals have been put forward to measure 
the polarization of the gluon among which is the proposal to measure angular 
correlation effects in the splitting process of a polarized gluon into a 
pair of gluons or quarks~\cite{gluon5}. Latter proposal has led to a 
beautiful confirmation of the presence of the three-gluon vertex in the 
$e^+e^-$ data~\cite{gluon6} (see also~\cite{gluon7}). 

The earlier calculations of the gluon's polarization in 
$e^+e^-$ annihilations had been done for massless fermions which was quite 
sufficient for the purposes of that period~\cite{gluon1,gluon2}. In the 
meantime the situation has changed in so far as the heavy top quark has 
been discovered whose production properties in $e^+e^-$ annihilations will 
be studied in the proposed Next-Linear-Collider (NLC). Typical running 
energies of the NLC would extend from $t\bar t$-threshold at about $350\GeV$ 
to maximal energies of about $550\GeV$. It is quite clear that top mass 
effects cannot be neglected in this energy range even at the highest 
c.m.~energies. It is therefore timely to redo the calculations 
of~\cite{gluon1,gluon2} for heavy quarks and to investigate the influence of 
heavy quark mass effects on the polarization observables of the gluon.

As is usual we shall represent the two-by-two differential density matrix 
$d\ssigma=d\sigma_{\lambda_G\lambda'_G}$ of the gluon with gluon helicities 
$\lambda_G=\pm 1$ in terms of its components along the unit matrix and the 
three Pauli matrices. Accordingly one has
\begin{equation}\label{eqn1}
d\ssigma=\frac12(d\sigma\oone+d\sigma^x\ssigma_x+d\sigma^y\ssigma_y
  +d\sigma^z\ssigma_z)
\end{equation}
where $d\sigma$ is the unpolarized differential rate and 
$d\vec\sigma=(d\sigma^x,d\sigma^y,d\sigma^z)$ are the three components of 
the (unnormalized) differential Stokes vector.

Specifying to $e^+e^-\rightarrow q(p_1)\bar q(p_2)G(p_3)$ we perform an 
azimuthal and polar averaging over the relative beam-event orientation. 
After azimuthal averaging the $y$-component of the Stokes vektor $d\sigma^y$ 
drops out~\cite{gluon1,gluon2}. One retains only the $x$- and 
$z$-components of the Stokes vector which are referred to as the gluon's 
linear polarization in the event plane and the circular polarization of 
the gluon, respectively. In this report we study the differential energy 
distribution of the polarization of the gluon, differential with regard to 
the scaled gluon energy $x=2p_3\cdot q/q^2$. After having integrated over 
the quark (or antiquark) energy the circular polarization of the gluon 
averages to zero due to $CP$-invariance\footnote{A nonvanishing circular 
polarization component is retained if one applies flavour tagging on the 
quark/antiquark. Even then, the circular polarization of the gluon turns 
out to be quite small.}. The differential unpolarized and polarized rates 
(with $q=p_1+p_2+p_3$) are then given by
\begin{equation}\label{eqn2}
\frac{d\sigma^{(x)}}{dx}
  =g_{11}\frac{d\sigma_{U+L}^{1(x)}}{dx}
  +g_{12}\frac{d\sigma_{U+L}^{2(x)}}{dx}.
\end{equation}
The notation $d\sigma^{(x)}$ stands for either $d\sigma$ or $d\sigma^x$, and 
the same for $d\sigma_{U+L}^{i(x)}$ (the index $i$ is explained later on). 
This notation closely follows the one in~\cite{gluon8} where the 
nomenclature $(U+L)$ has been used to denote the total rate 
($U$: unpolarized transverse, $L$: longitudinal).

The electro-weak cross section is written in modular form in terms of two 
building blocks. The first building block specifies the electro-weak model 
dependence through the parameters $g_{ij}$ ($i,j=1,\ldots, 4$). They are 
given by
\begin{eqnarray}\label{eqn3}
g_{11}&=&Q_f^2-2Q_fv_ev_f\real\chi_Z+(v_e^2+a_e^2)(v_f^2+a_f^2)|\chi_Z|^2,
  \nonumber\\
g_{12}&=&Q_f^2-2Q_fv_ev_f\real\chi_Z+(v_e^2+a_e^2)(v_f^2-a_f^2)|\chi_Z|^2,\\
g_{41}&=&2Q_fa_ev_f\real\chi_Z-2v_ea_e(v_f^2+a_f^2)|\chi_Z|^2,\nonumber\\
g_{42}&=&2Q_fa_ev_f\real\chi_Z-2v_ea_e(v_f^2-a_f^2)|\chi_Z|^2 \nonumber
\end{eqnarray}
where, in the Standard Model, 
$\chi_Z(q^2)=gM_Z^2q^2/(q^2-M_Z^2+iM_Z\Gamma_Z)^{-1}$, with $M_Z$ and 
$\Gamma_Z$ the mass and width of the $Z^0$ and 
$g=G_F(8\sqrt2\pi\alpha)^{-1}\approx 4.49\cdot 10^{-5}\GeV^{-2}$. $Q_f$ are 
the charges of the final state quarks to which the electro-weak currents 
directly couple; $v_e$ and $a_e$, $v_f$ and $a_f$ are the electro-weak 
vector and axial vector coupling constants. For example, in the 
Weinberg-Salam model, one has $v_e=-1+4\sin^2\theta_W$, $a_e=-1$ for 
leptons, $v_f=1-\frac83\sin^2\theta_W$, $a_f=1$ for up-type quarks 
($Q_f=\frac23$), and $v_f=-1+\frac43\sin^2\theta_W$, $a_f=-1$ for down-type 
quarks ($Q_f=-\frac13$). In this paper we use Standard Model couplings
with $\sin^2\theta_W=0.226$.

The second building block is determined  by the hadron dynamics, i.e.\ 
by the current-induced production of a heavy quark pair with subsequent 
gluon emission. We shall work in terms of the components of the polarized 
and unpolarized hadronic tensor $H_{U+L}^{i(x)}$ which are related to the 
differential rate by
\begin{equation}
\frac{d\sigma_{U+L}^{i(x)}}{dx}
  =\frac{\alpha^2}{24\pi q^2}H_{U+L}^{i(x)}(x).
\end{equation}
The index $i=1,2$ specifies the current composition in terms of the two 
parity-conserving products of the vector and the axial vector currents 
according to (we drop all further indices on the hadron tensor) 
\begin{equation}
H^1=\frac12(H^{VV}+H^{AA}),\quad H^2=\frac12(H^{VV}-H^{AA}).
\end{equation}

Eq.~(\ref{eqn2}) gives the differential cross section for unpolarized beams. 
The case of longitudinally polarized beams can easily be included and leads 
to the replacement
\begin{equation}
g_{1i}\rightarrow(1-h^-h^+)g_{1i}+(h^--h^+)g_{4i}\qquad(i=1,2)
\end{equation}
in the unpolarized and linearly polarized differential rates 
$d\sigma_{U+L}^{i(x)}$ where the electroweak coefficients $g_{4i}$ are given 
in Eqs.~(\ref{eqn3}) and where $h^-$ ($h^+$) denote the longitudinal 
polarization of the electron (positron). 

The various pieces of the hadronic tensor can be calculated from the 
relevant Feynman diagrams. After integration over the quark (or antiquark) 
energy one obtains
\begin{eqnarray}\label{eqn4}
H_{U+L}^1(x)&=&N\bigg[-2\left(\frac{4-\xi}x-(4-\xi)+2x\right)\frac1xw_+(x)
  \nonumber\\&&+\left(\frac{(4-\xi)(2-\xi)}x-2(4-\xi)+(4+\xi)x\right)
  t_{\ell+}(x)\bigg],\nonumber\\
\tilde H_{U+L}^2(x)&=&\xi N\bigg[-6\left(\frac1x-1\right)\frac1xw_+(x) 
  +\left(3\frac{2-\xi}x-6-x\right)t_{\ell+}(x)\bigg],\nonumber\\
\tilde H_{U+L}^{1x}(x)&=&(4-\xi)N\bigg[-2\left(\frac1x-1\right)\frac1xw_+(x) 
  +\left(\frac{2-\xi}x-2\right)t_{\ell+}(x)\bigg],\nonumber\\
\tilde H_{U+L}^{2x}(x)&=&3\xi N\bigg[-2\left(\frac1x-1\right)\frac1xw_+(x) 
  +\left(\frac{2-\xi}x-2\right)t_{\ell+}(x)\bigg]
\end{eqnarray}
where $\xi=4m_q^2/q^2$, $v=\sqrt{1-\xi}$, $N=2\alpha_sN_CC_Fq^2/\pi v$ with 
$N_C=3$ and $C_F=4/3$,
\begin{equation}
w_+(x)=x\sqrt{\frac{1-x-\xi}{1-x}}
\end{equation}
and
\begin{equation}
t_{\ell+}(x)=\ln\left(\frac{\sqrt{1-x}+\sqrt{1-x-\xi}}
  {\sqrt{1-x}-\sqrt{1-x-\xi}}\right).
\end{equation}

It is not difficult to recover the mass zero result from Eqs.~(\ref{eqn4}). 
Taking the $\xi\rightarrow 0$ limit in Eqs.~(\ref{eqn4}) one has 
\begin{eqnarray}\label{eqn5}
H_{U+L}^1(x)&\rightarrow&-32N\left(\frac2x-2+x\right)\ln\xi,\nonumber\\
H_{U+L}^{1x}(x)&\rightarrow&-32N\left(\frac2x-2\right)\ln\xi 
\end{eqnarray}
and $H_{U+L}^{2}(x)$, $H_{U+L}^{2x}(x)\to 0$.

The normalized linear polarization $P^{x}(x)$ of the gluon is given by 
the  normalized Stokes vector components. One has
\begin{equation}
P^{x}(x)=\frac{d\sigma^{x}/dx}{d\sigma/dx}
\end{equation} 
In Fig.~1(a) we plot the linear polarization of the gluon as a function of 
the gluon's fractional energy $x/x_{\rm max}$ for the top and charm quark 
cases ($x_{\rm max}= 1-\xi$). We use a c.m.~energy of $500\GeV$. At both 
ends of the spectrum the linear polarization of the gluon is fixed by 
general and model independent considerations. At the soft gluon end it is 
well-known that the linear polarization is 100\% while at the hard end of 
the spectrum the linear polarization has to go to zero for the simple reason 
that one can no longer define a hadronic plane in this collinear 
configuration. These limits can be easily verified by taking the corresponding 
$x\rightarrow 0$ and $x\rightarrow 1-\xi$ limits in Eqs.~(\ref{eqn4}). We 
have chosen to compare the polarization of the gluon in the top and charm 
quark cases at the same fractional energy $E_G/E_G^{\rm max}=x/x_{\rm max}$. 
For a given fractional energy $x/x_{\rm max}$ the linear polarization of the 
gluon is always higher in the top quark case than in the charm quark or mass 
zero case. Contrary to this one finds a higher degree of polarization in the 
charm quark case than in the top quark case when comparing the linear 
polarization at {\it fixed gluon energies}. However, a comparison at a 
fixed fractional energy of the gluon is more appropiate from a physics 
point of view in particular if one is interested in the average linear 
polarization of the gluon to be discussed later on. The linear polarization 
of the gluon remains above 50\% for 85\% of the avaliable energy range in 
the top quark case. As Fig.~1(b) shows, the rate for top quark production 
is strongly weighted towards smaller gluon energies where the linear 
polarization is large. We anticipate a large average linear polarization of 
the top quark. In the charm quark case the linear gluon polarization is 
already quite close to the zero mass case which, according to 
Eq.~(\ref{eqn5}), is given by\footnote{The limiting value of the linear 
polarization agrees with the corresponding result Eqs.~(4--9) 
in~\cite{gluon1} (second reference) in the limit $x_0,\delta\rightarrow 0$.}
\begin{equation}\label{eqn6}
P^x(x)=\frac{1-x}{1-x+\frac12x^2}.
\end{equation}
The good quality of the zero mass formula Eq.~(\ref{eqn6}) when applied to 
the charm quark case ($\sqrt{q^2}=500\GeV$ and $m_c=1.3\GeV$) must be judged 
against the fact that the dominating logarithmic term $\log\xi=-10.52$ is 
not yet overly large.

The linear polarization is flavour independent (and beam polarization 
independent) in the zero mass limit since the flavour dependent $g_{11}$ 
factor drops out in the ratio Eq.~(11). This is different in the massive 
case where flavour dependence comes in through the nonvanishing of the 
hadron tensor component $H^2=\frac12(H^{VV}-H^{AA})$. It can, however, be 
checked that the dependence on the electro-weak parameters is also quite 
weak in the massive quark case. The reason is two-fold. First the 
contributions of $H^2$ and $H^{2x}$ are somewhat suppressed even for top 
quark pair production. Secondly the rate is dominated by one photon 
exchange at the energy $\sqrt{q^2}=500\GeV$ leading again to a near 
cancellation of the electro-weak model dependence.

The last step is the integration over the second phase-space parameter $x$.
It is clear that we have to introduce a gluon energy cut-off at the soft 
end of the gluon spectrum in order to keep the rate finite. Denoting the 
cut-off energy by $E_c=\lambda\sqrt{q^2}$ the integration extends from 
$x=2\lambda=2E_c/\sqrt{q^2}$ to $x=1-\xi$. We obtain
\begin{eqnarray}
H_{U+L}^1&=&N\bigg[-(4-\xi)\Big[2\GG(-1)-2\GG(0)-(2-\xi)\GL(-1)+2\GL(0)\Big]
  \nonumber\\&&\qquad\qquad-4\GG(1)+(4+\xi)\GL(1)\bigg],\nonumber\\
H_{U+L}^2&=&\xi N\bigg[-3\Big[2\GG(-1)-2\GG(0)-(2-\xi)\GL(-1)+2\GL(0)\Big]
  -\GL(1)\bigg],\nonumber\\
H_{U+L}^{1x}&=&-(4-\xi)N\bigg[2\GG(-1)-2\GG(0)-(2-\xi)\GL(-1)
  +2\GL(0)\bigg],\nonumber\\
H_{U+L}^{2x}&=&-3\xi N\bigg[2\GG(-1)-2\GG(0)-(2-\xi)\GL(-1)
  +2\GL(0)\bigg]
\end{eqnarray}
using the integrals
\begin{eqnarray}
\GG(m)&:=&\int_{2\lambda}^{1-\xi}x^{m-1}w_+(x)dx,\\[12pt]
\GG(-1)&=&-\ln\left(\frac{1+A}{1-A}\right)
  -v\ln\left(\frac{v-A}{v+A}\right),\\[12pt]
\GG(0)&=&\frac{\xi A}{1-A^2}-\frac\xi2\ln\left(\frac{1+A}{1-A}\right),\\[12pt]
\GG(1)&=&\frac{\xi A}{4(1-A^2)^2}(4-\xi-(4+\xi)A^2)
  -\frac\xi8(4-\xi)\ln\left(\frac{1+A}{1-A}\right),\\[12pt]
\GL(m)&:=&\int_{2\lambda}^{1-\xi}x^mt_{\ell+}(x)dx,\\[12pt]
\GL(-1)&=&\frac12\ln\left(\frac{1-A^2}4\right)\ln\left(\frac{1+A}{1-A}\right)
  -\ln\left(\frac{1+v}{1-v}\right)\ln\left(\frac{v-A}{v+A}\right)\nonumber\\&&
  +\Li\left(\frac{1+A}{2}\right)-\Li\left(\frac{1-A}2\right)
  +\Li\left(-\frac{v+A}{1-v}\right)\\&&
  -\Li\left(-\frac{v-A}{1-v}\right)+\Li\left(\frac{v-A}{1+v}\right)
  -\Li\left(\frac{v+A}{1+v}\right),\nonumber\\[12pt]
\GL(0)&=&\frac\xi2\left(\frac{1+A^2}{1-A^2}\right)
  \ln\left(\frac{1+A}{1-A}\right)-\frac{\xi A}{1-A^2},\\[12pt]
\GL(-1)&=&\frac\xi{16(1-A^2)^2}(8-5\xi-6\xi A^2-(8-3\xi)A^4)
  \ln\left(\frac{1+A}{1-A}\right)\nonumber\\&&
  +\frac{\xi A}{8(1-A^2)^2}(-8+5\xi+(8-3\xi)A^2)
\end{eqnarray}
where
\begin{equation}
A=\sqrt{\frac{1-2\lambda-\xi}{1-2\lambda}}.
\end{equation}

In Fig.~2 we show a plot of the average linear polarization of the gluon 
as a function of the c.m.~energy $\sqrt{q^2}$ for three different cut-off 
values $E_c=\lambda\sqrt{q^2}=5$, $10$ and $15\GeV$. Gluon energies of this 
magnitude are sufficient to make the corresponding gluon jets detectable. 
Because of the ``dead cone'' effect in the massive case, the gluon jet would 
be pointing away from the original top or antitop direction. The average 
linear polarization of the gluon rises steeply from threshold and quickly 
attains very high values around $95\%$ for the top quark case. In the charm 
quark case the average linear polarization is also large but is somewhat 
smaller than in the top quark case. The linear polarization $P^x$ becomes 
larger for smaller values of $E_c=\lambda\sqrt{q^2}$ and tends to one as 
$\lambda$ goes to zero. The approach to the asymptotic value $P^x=1$ is, 
however, rather slow. 

In the leading log aproximation as $\lambda\rightarrow 0$ the linear 
polarization formula considerably simplifies. The leading log contributions 
can be easily identified in the terms $\GG(-1)$ and $\GL(-1)$. They are 
obtained by setting
\begin{equation}
\ln\left(\frac{v-A}{v+A}\right)\rightarrow 
\ln\left(\frac{\lambda \xi}{2 v^2}\right)\qquad\mbox{for }
  \lambda\rightarrow 0.
\end{equation}
In the other terms one can savely set $A=v$ as $\lambda\rightarrow 0$. We 
mention that the leading log representation of the linear polarization gives 
very accuarate numerical results for the above range of cut-off values except 
for energies close to theshold. For example, for top production and for 
$E_c=\lambda\sqrt{q^2}=15\GeV$ the leading log result is $0.13\%$ below the 
full result at $\sqrt{q^2}=500\GeV$ and $0.12\%$ at $\sqrt{q^2}=1000\GeV$.
 
In conclusion we have computed gluon polarization effects in the process 
$e^+e^-\rightarrow t\bar tG$. Compared to the zero quark mass case the 
average linear polarization of the gluon is somewhat enhanced through quark 
mass effects. If one aims to study gluon polarization effects 
in the splitting process $e^+e^-\rightarrow t\bar tG(\rightarrow GG, q\bar q)$ 
the present on-shell calcuation should be sufficient to identify and 
discuss the leading effects of gluon polarization without that one has to 
perform a full $O(\alpha^2_s)$ calculation of
$e^+e^-\rightarrow t\bar tGG$ and 
$e^+e^-\rightarrow t\bar tq\bar q$~\cite{gluon5}.

\vspace{1truecm}\noindent
{\bf Acknowledgements:} This work is partially supported by the BMBF, FRG, 
under contract No.\ 06MZ865, and by HUCAM, EU, under contract No.\ 
CHRX-CT94-0579. S.G. acknowledges financial support by the DFG, FRG. 
The work of J.A.L.\ is supported by the DAAD, FRG.

\vspace{1cm}
\centerline{\Large\bf Figure Captions}
\vspace{.5cm}
\newcounter{fig}
\begin{list}{\bf\rm Fig.\ \arabic{fig}:}{\usecounter{fig}
\labelwidth1.6cm\leftmargin2.5cm\labelsep.4cm\itemsep0ex plus.2ex}
\item a) Energy dependence of the linear polarization of the gluon.\\
  b) Energy dependence of the differential cross section
  $e^+e^-\rightarrow t\bar tG$ and $e^+e^-\rightarrow c\bar cG$
\item Average linear polarization of the gluon in 
  $e^+e^-\rightarrow t\bar tG$ and $e^+e^-\rightarrow c\bar cG$ 
  for different values of the cut-off energy $E_c$ as function of 
  the c.m.~energy $\sqrt{q^2}$
\end{list}
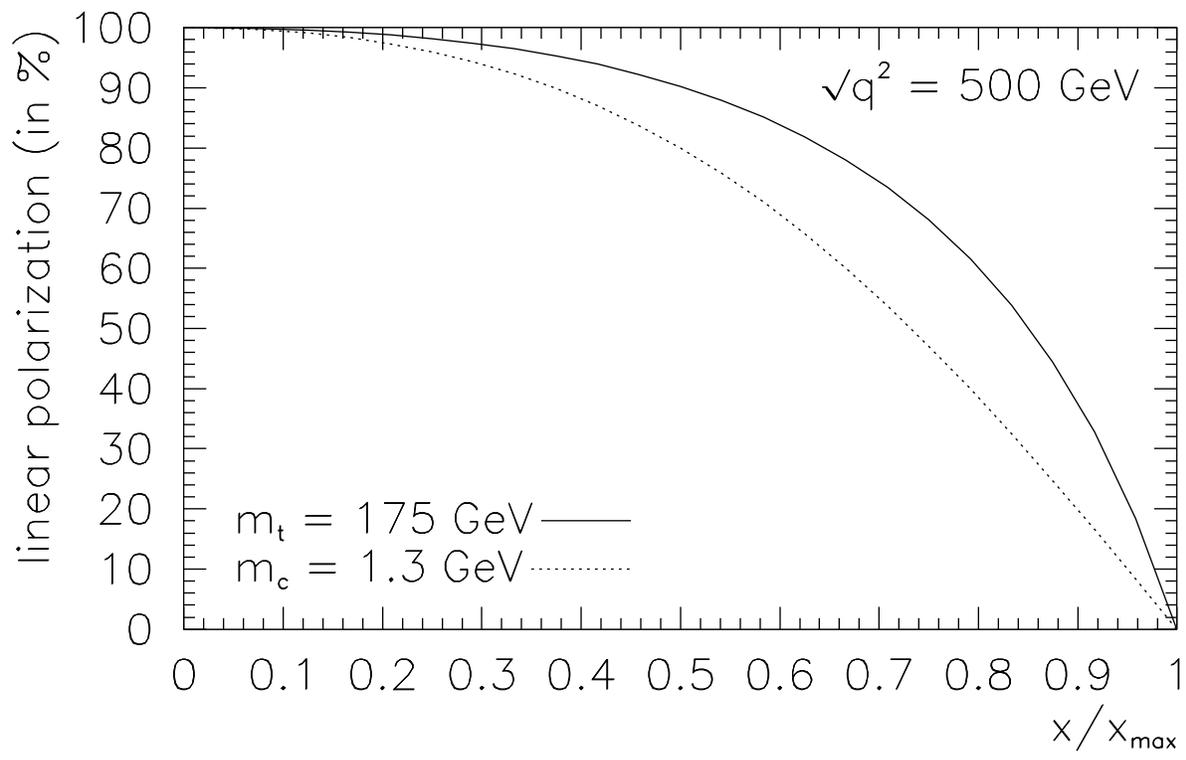
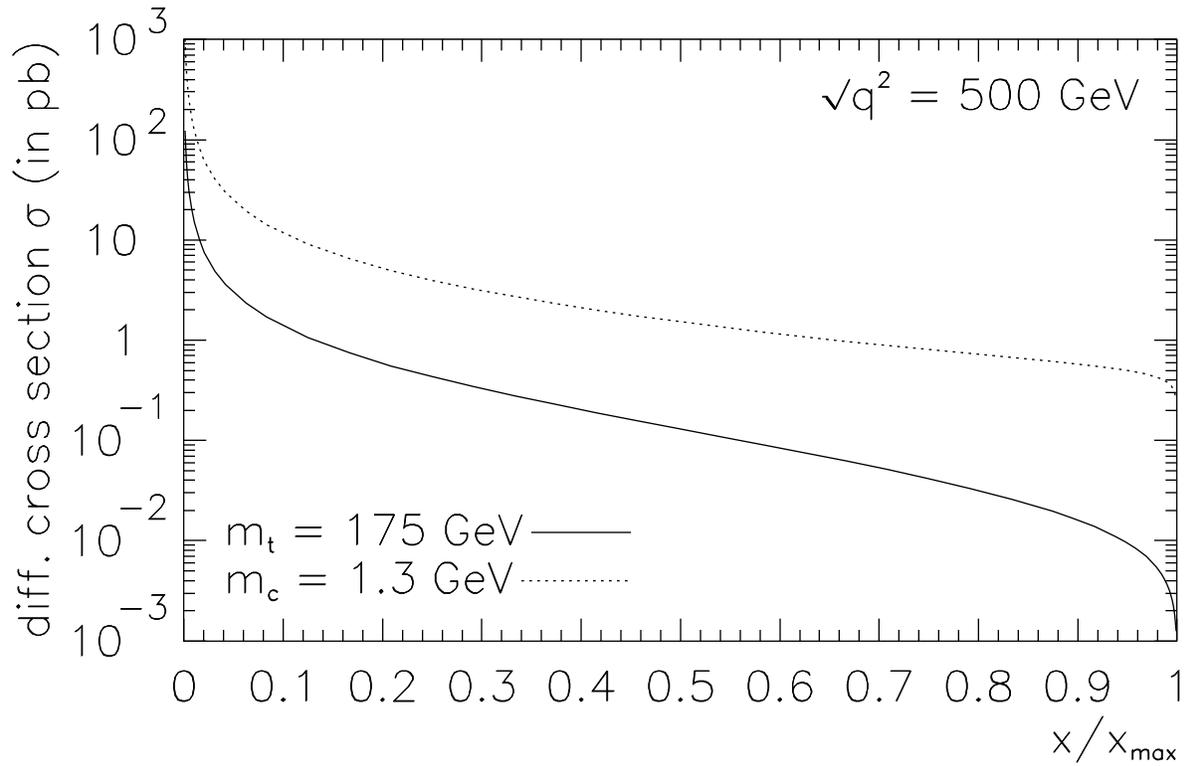
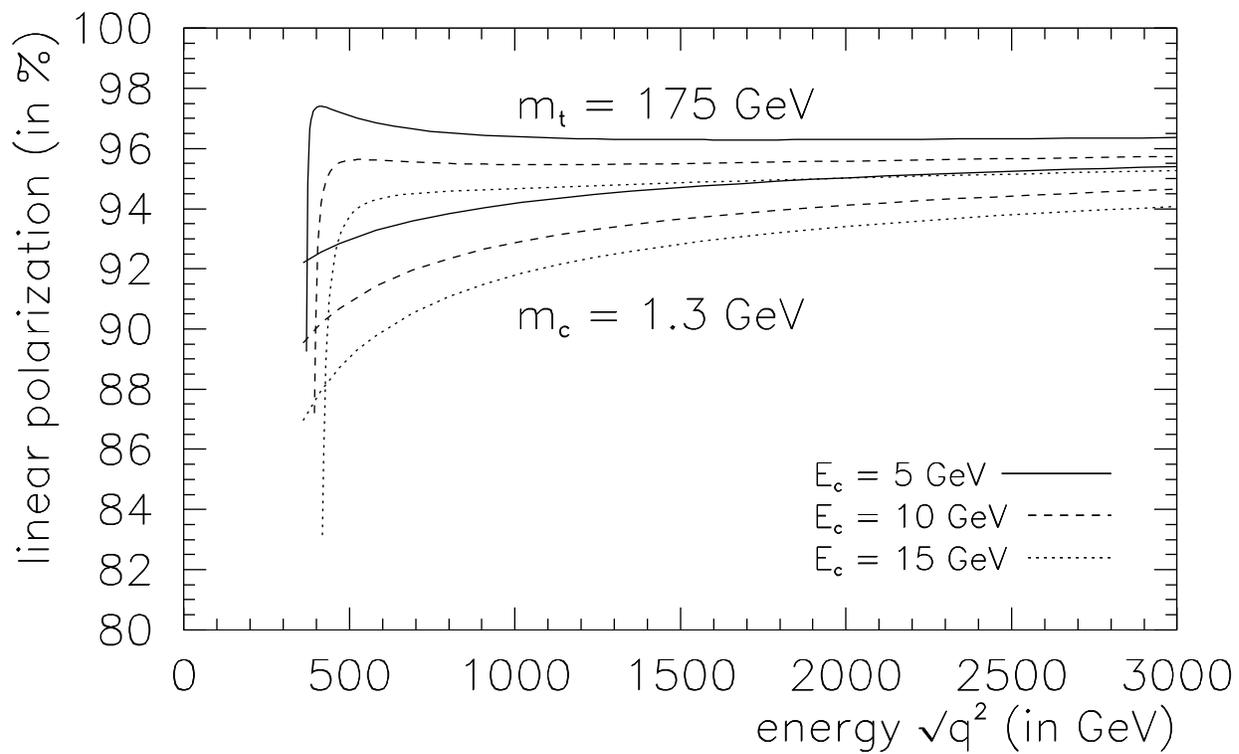
\end{document}